\documentclass[12pt]{iopart}

\usepackage{graphicx}

\newcommand{\st}{_\mathrm{st}}                      
\newcommand{\abs}[1]{|#1|}                          
\newcommand{\avg}[1]{\langle #1 \rangle}            
\newcommand{\avgst}[1]{\langle #1 \rangle\st}       
\newcommand{\hg}[2]{{}_{#1}F_{#2}}                  

\begin{document}

\title[Exact solution of DMFT for a linear system with annealed disorder]{Exact solution of Dynamical Mean-Field Theory for a linear system with annealed disorder}

\author{Francesco Ferraro$^{*,1,2,3}$, Christian Grilletta$^{*,1}$, Amos Maritan$^{1,2,3}$, Samir Suweis$^{1,2,4}$ and Sandro Azaele$^{1,2,3}$}

\address{$^1$ Laboratory of Interdisciplinary Physics, Department of Physics and Astronomy\\``G. Galilei'', University of Padova, Padova, Italy}
\address{$^2$ INFN, Sezione di Padova, via Marzolo 8, Padova, Italy}
\address{$^3$ National Biodiversity Future Center, Piazza Marina 61, 90133 Palermo, Italy}
\address{$^4$ Padova Neuroscience Center, University of Padova, Padova, Italy}
\address{$^*$ These authors contributed equally}

\eads{
    \mailto{francesco.ferraro.4@phd.unipd.it},
    \mailto{christian.grilletta@unipd.it}
}

\begin{abstract}
    We investigate a disordered multi-dimensional linear system in which the interaction parameters vary stochastically in time with defined temporal correlations. We refer to this type of disorder as ``annealed'', in contrast to quenched disorder in which couplings are fixed in time. We extend Dynamical Mean-Field Theory to accommodate annealed disorder and employ it to find the exact solution of the linear model in the limit of a large number of degrees of freedom. Our analysis yields analytical results for the non-stationary auto-correlation, the stationary variance, the power spectral density, and the phase diagram of the model. Interestingly, some unexpected features emerge upon changing the correlation time of the interactions. The stationary variance of the system and the critical variance of the disorder are generally found to be a non-monotonic function of the correlation time of the interactions. We also find that in some cases a re-entrant phase transition takes place when this correlation time is varied.
\end{abstract}

\section{Introduction}

Large systems of coupled differential equations have attracted great interest for their applicability in neural networks \cite{sompolinsky1988chaos}, ecology \cite{bunin2017ecological}, theoretical biology \cite{opper1992phase}, and many other fields \cite{anand2009stability,baron2021consensus,godreche2018characterising}. The parameters specifying the interactions between the degrees of freedom composing such systems are prohibitively numerous and possibly altogether unknowable. A sensible approach to deal with both limitations is to replace these parameters with numbers drawn from random distributions characterized by fewer parameters, which effectively lowers the dimensionality of the model and allows analytical progress. This approach has a long standing history in physics, starting from the study of nuclear structure \cite{wigner1967random}, but has also been successfully applied to ecology \cite{allesina2012stability} and to complex systems in general \cite{may1972will}.

However, in real systems, the assumption that interactions in these models are fixed in time is not always satisfied. For instance, synaptic plasticity in neuronal populations consists of long-term potentiation or depression, resulting in temporal fluctuations in the interactions strength between neurons \cite{sjostrom2001rate, magee2020synaptic}. In an ecological setting, it is known that the strength of interactions between species may fluctuate on a timescale comparable to that of population dynamics \cite{suweis2013emergence,fiegna2015evolution,ushio2018fluctuating}.

For this reason, a recent work \cite{suweis2023generalized} has studied a central model in theoretical ecology, the Generalized Lotka-Volterra (GLV) equations, taking into account the time dependence of the interactions between species. In this work, interactions were modeled as colored noise with a characteristic correlation time $\tau$, unlike previous approaches \cite{bunin2017ecological,galla2018dynamically,biroli2018marginally} with random quenched couplings. We dub this type of random interactions \textit{annealed disorder}. This simple modification of the GLV equations however made exact results beyond reach, except for the limiting cases of $\tau\to0$ and $\tau\to\infty$.

Here, we tackle the problem of finding the exact solution of the simpler case of a linear interacting system with annealed disorder. We solve this model in the limit of a large number of degrees of freedom employing Dynamical Mean-Field Theory (DMFT). This is a tool routinely used in the study of non-equilibrium disordered systems \cite{cugliandolo2023recent}, which is tailored here for the case of annealed interactions. Our work thus demonstrates the first exact solution of DMFT in the case of time-correlated stochastic interactions.

We show that DMFT leads to a stochastic differential equation that is formally equivalent to that of a one-dimensional Ornstein-Uhlenbeck process, with the difference that the DMFT process is driven by a noise which is self-consistent and colored. The DMFT process is Gaussian and its first two moments can be exactly determined. We are thus able to obtain analytical results for the non-stationary autocorrelation, the stationary variance, and the power spectral density. At odds with the one-dimensional Ornstein-Uhlenbeck process, driven by white noise, the DMFT process displays a non-trivial phase diagram. A stationary state is not reached if the mean or the variance of the random interactions exceed some critical value, which we are also able to determine analytically.

Despite the simplicity of the model considered here, interesting features arise as the correlation time of the couplings changes. For example, the stationary variance of the process and the critical variance of the disorder are generally found to be a non-monotonic function of the correlation time of the interactions $\tau$. We also find that in some cases a re-entrant phase transition takes place in $\tau$.

The paper is organized as follows: in \sref{sec:model} we present the model; in \sref{sec:DMFT} we derive its DMFT equation, which is solved in \sref{sec:DMFT-solution}; in \sref{sec:phase-diagram} we derive the phase diagram of the model; in \sref{sec:white-quenched} simplified results in the limit of white-noise and quenched disorder are given; in \sref{sec:UCNA} we compare the exact solution with an approximation; we conclude with a summary of the results and possibilities for further future research directions.

\section{Model}
\label{sec:model}

\begin{figure}
    \centering
    \includegraphics[width=\textwidth]{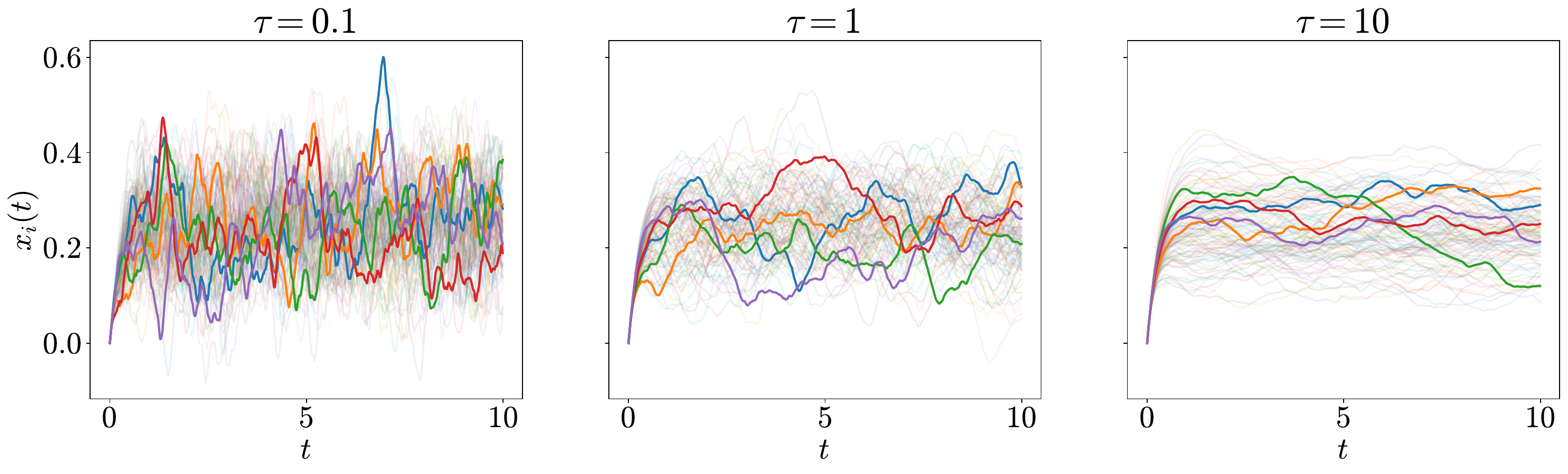}
    \caption{Example of trajectories of the linear system \eref{eq:linear-equation} for different values of the correlation time $\tau$ of the annealed disorder. In each plot five trajectories chosen at random are highlighted. The model has been simulated with $N=1000$, $dt=0.01$, $h=1$, $k=4$, $\mu=0$, $\sigma=1$, and initial condition $x_i(0)=0$ for all $i$.}
    \label{fig:example-trajs}
\end{figure}

We consider a system with $N$ degrees of freedom $\{x_i(t)\}$, $i=1,\dots,N$, that interact linearly
\begin{equation}
    \dot{x}_i(t) = h - k x_i(t) + \sum_{j \neq i} \alpha_{ij}(t) x_j(t).
    \label{eq:linear-equation}
\end{equation}
Here $h$ and $k$ are fixed parameters and the annealed disorder $\alpha_{ij}(t)$ is Gaussian colored noise. Explicitly, it is given by
\begin{equation}
    \alpha_{ij}(t) = \frac{\mu}{N} + \frac{\sigma}{\sqrt{N}}z_{ij}(t),
\end{equation}
where $\mu$ and $\sigma$ are fixed parameters, and $z_{ij}(t)$ are independent Ornstein-Uhlenbeck processes with
\begin{eqnarray}
    \avg{z_{ij}(t)} = 0, \\
    \avg{z_{ij}(t)z_{ij}(t')} = Q(t-t'),
\end{eqnarray}
where
\begin{eqnarray}
    Q(t) = \frac{1+2\tau/\tau_0}{2\tau} \exp(-\abs{t}/\tau).
    \label{eq:noise-amplitude}
\end{eqnarray}
The scaling of $\alpha_{ij}(t)$ with the number of degrees of freedom $N$ is chosen to have a non-trivial $N\to\infty$ limit. The noise amplitude in \eref{eq:noise-amplitude} is chosen so that in the limit of $\tau\to0$ the annealed disorder becomes white noise, while in the limit $\tau\to\infty$ we recover the case of quenched disorder. We will also set $\tau_0=1$ without loss of generality. \Fref{fig:example-trajs} shows examples of the dynamics of the system for different correlation times $\tau$.

The general case of disordered or different $h_i$, $k_i$ for each $i$, possibly time-dependent, is a possible extension of our framework, but is not discussed in this work.

\section{Dynamical Mean-Field Theory}\label{sec:DMFT}
In order to solve the linear system \eref{eq:linear-equation} we employ Dynamical Mean-Field Theory. This tool, which has its origin in the study of non-equilibrium disordered systems, replaces a fully-connected $N$-body problem with a mean-field, single-body problem. The difficulty in treating the original problem appears through the presence of self-consistency relations, as is the case for all mean-field approaches. DMFT is exact in the limit $N\to\infty$.

The DMFT equation can be derived in the case of quenched interactions for a large class of models using the generating functional formalism \cite{galla2018dynamically} or the dynamical cavity method \cite{roy2019numerical}. In the case where interactions are not quenched, but vary stochastically, the DMFT equation has been derived in the context of the GLV equations through a simple extension of the quenched case \cite{suweis2023generalized}.

The DMFT equation relative to the linear system \eref{eq:linear-equation}, which is derived in \ref{appendix:DMFT_derivation} along the lines of \cite{suweis2023generalized}, is the following stochastic differential equation for a representative degree of freedom of the system $x(t)$
\begin{equation}
    \dot{x}(t) = h - kx(t) + \mu \avg{x(t)} + \sigma \eta(t).
    \label{eq:DMFT}
\end{equation}
In this equation $\eta(t)$ is a colored, non-stationary, Gaussian noise with mean and correlations given self-consistently by
\begin{eqnarray}
    \avg{\eta(t)} = 0, \\
    \avg{\eta(t)\eta(t')} = Q(t-t') \avg{x(t)x(t')},
\end{eqnarray}
where the averages $\avg{x(t)}$ and $\avg{x(t)x(t')}$ are understood to be over solutions of \eref{eq:DMFT}.

\Eref{eq:DMFT} thus defines a stochastic process which is both self-consistent and colored. As shown in the following, this makes the DMFT process fundamentally different from a one-dimensional Ornstein-Uhlenbeck process. For example, the DMFT process displays a non-trivial phase diagram, in contrast with that of the Ornstein-Uhlenbeck, which invariably settles into a stationary state.

\section{Exact solution of Dynamical Mean-Field Theory}\label{sec:DMFT-solution}
To solve the DMFT equation we start by noticing that the formal solution of \eref{eq:DMFT} is
\begin{equation}
    x(t) = x_0 \rme^{-kt} + \int_0^t \rmd s \rme^{-k(t-s)}\left[h + \mu \avg{x(s)} + \sigma\eta(s)\right],
    \label{eq:DMFT-solution}
\end{equation}
assuming a fixed initial condition $x_0$ at $t=0$. Extending the results of this work to the general case of an initial condition drawn from a probability distribution is possible, but not discussed here. 

\Eref{eq:DMFT-solution} shows that the process $x(t)$, being a linear combination of the Gaussian noise $\eta(t)$ at different times, is a Gaussian process. The process $x(t)$ is therefore completely specified by its mean $\avg{x(t)}$ and autocorrelation $C(t,t') = \avg{x(t)x(t')} - \avg{x(t)}\avg{x(t')}$.  In the rest of this section we derive and solve closed equations for the mean and autocorrelation, thus exactly solving the DMFT equation \eref{eq:DMFT}.

We also notice that, since $x(t)$ is a Gaussian process with a fixed initial condition, its single-time probability distribution is given by a Gaussian
\begin{equation}
    P(x,t) = \frac{1}{\sqrt{2 \pi C(t, t) }} \exp{\left[-\frac{\left(x-\langle x(t) \rangle \right)^2 }{2 C(t, t) }  \right]}.
\label{eq:P(xt)}
\end{equation}

\subsection{Mean}

The time evolution of the mean is simply found by taking the average of \eref{eq:DMFT}
\begin{equation}
    \frac{\rmd\avg{x(t)}}{\rmd t} = h - (k-\mu) \avg{x(t)},
\end{equation}
which has solution
\begin{equation}
    \avg{x(t)} = x_0 \rme^{-(k-\mu)t} + \frac{h}{k-\mu}\left[1-\rme^{-(k-\mu)t}\right].
    \label{eq:M(t)-solution}
\end{equation}
For the stationary state of the system to be reached we have the necessary condition
\begin{equation}
    \mu < k,
\end{equation}
and when the stationary state is reached its mean is given by
\begin{equation}
    \avgst{x} = \frac{h}{k-\mu}.
    \label{eq:avgeqx}
\end{equation}

\subsection{Autocorrelation}
To get a closed equation for the auto-correlation function we rewrite \eref{eq:DMFT} as
\begin{equation}
    \sigma\eta(t) = \dot{x} + k x - h - \mu \avg{x(t)}.
\end{equation}
By averaging the product $\sigma^2 \eta(t)\eta(t')$ and performing simple manipulations, assuming $t \neq t'$, we get the following partial derivative equation (PDE) for the autocorrelation
\begin{equation}
    \left[\partial_t\partial_{t'} + k(\partial_t + \partial_{t'}) + k^2 - \sigma^2 Q(t-t')\right] C(t,t') = f(t,t'),
    \label{eq:PDE}
\end{equation}
where
\begin{equation}
    f(t,t') = \sigma^2 Q(t-t') \avg{x(t)} \avg{x(t')}.
\end{equation}
This equation for $C(t,t')$ is valid only for $t \neq t'$, since it was derived with this assumption. As a consequence, one cannot set $t=t'$ to get a closed equation for the variance $C(t,t)$.

Since we are assuming the initial condition $x(0)=x_0$ to be fixed, the PDE has to be solved with the boundary conditions $C(t,0)=0$ and $C(0,t')=0$. The solution is found using the Riemann method in \ref{appendix:PDE-solution} and is
\begin{equation}
    C(t,t') = \rme^{-k(t+t')} \int_0^t \rmd s \int_0^{t'} \rmd s' \rme^{k(s+s')} A(s,s';t,t') f(s,s'),
    \label{eq:PDE-solution}
\end{equation}
where $A(s,s';t,t')$ is the Riemann function of the PDE, see \eref{eq:Riemann-function}.

\begin{figure}
    \centering
    \begin{minipage}{0.475\textwidth}
    \centering
        \includegraphics[width=\textwidth]{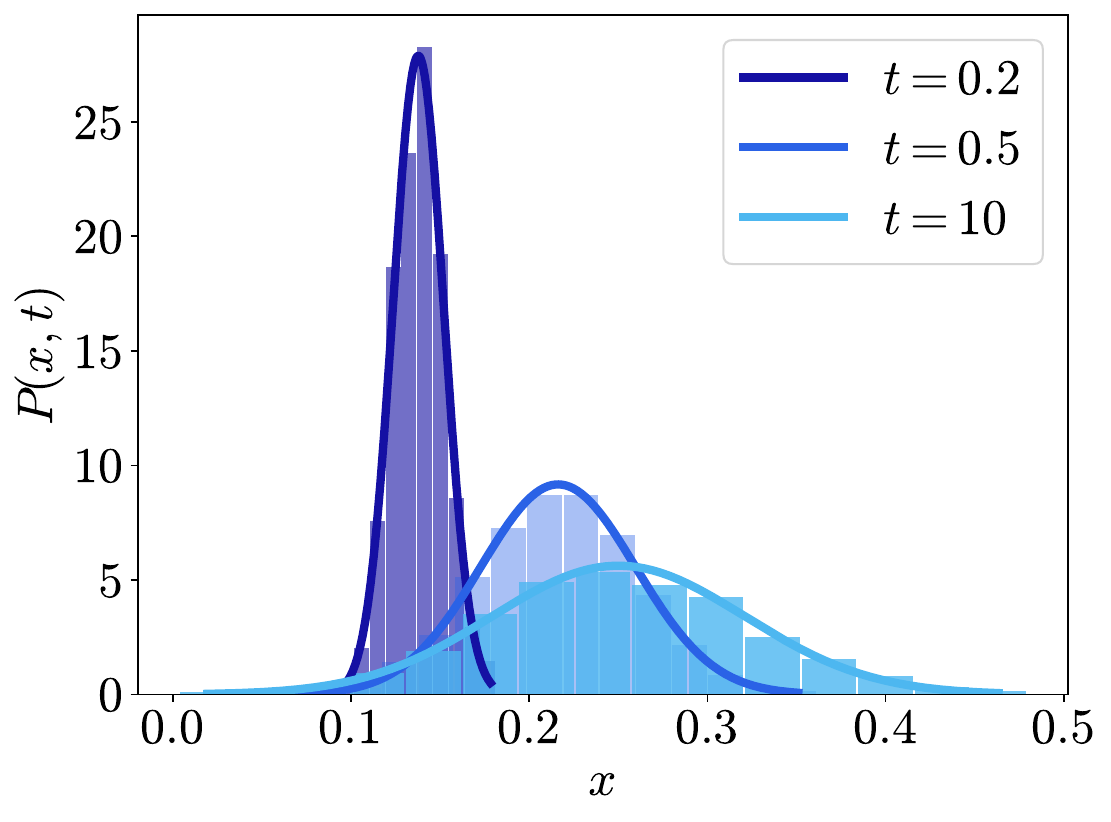}
    \end{minipage}\hfill%
    \begin{minipage}{0.475\textwidth}
    \centering
    \includegraphics[width=\textwidth]{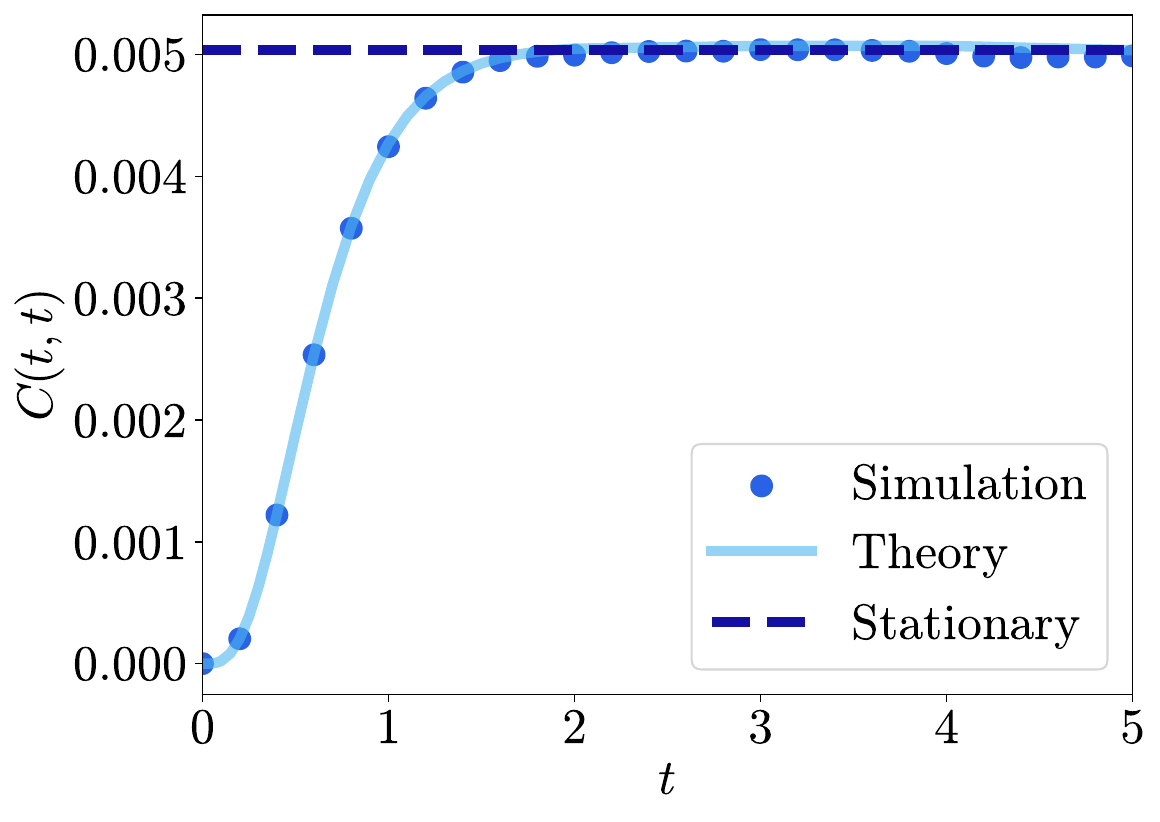}
    \end{minipage}
    \caption{Left: Comparison between the single-time probability distribution of $x(t)$ \eref{eq:P(xt)}, with mean and variance given by \eref{eq:M(t)-solution} and \eref{eq:PDE-solution}, and numerical simulations. The parameters are $N=1000$, $dt=0.01$, $h=1$, $k=4$, $\mu=0$, $\sigma=1$, and $\tau=1$. Right: Comparison between the time evolution of the variance of $x(t)$, found analytically by setting $t=t'$ in \eref{eq:PDE-solution}, and numerical simulations. The parameters are $N=1000$, $dt=0.01$, $h=1$, $k=4$, $\mu=0$, $\sigma=1$, and $\tau=1$. Each simulated point is the result of 100 iterations.}
\label{fig:figure2e3}
\end{figure}

Comparisons between this analytical result and those obtained by numerical integration of \eref{eq:linear-equation} are given in \fref{fig:figure2e3}. Even if derived only for $t \neq t'$, we observe numerically that \eref{eq:PDE-solution} gives the correct result also for $t=t'$, see \fref{fig:figure2e3}.

\subsection{Stationary autocorrelation and variance}
The process $x(t)$ is stationary for sufficiently long times, that is, $t\gg 1/(k-\mu)$, and its autocorrelation depends only on the difference $t-t'$. Thus, in this case $C(t,t') = C\st(t-t')$ and at stationarity \eref{eq:PDE} reduces to an ODE
\begin{equation}
    -C\st''(t) + \left[k^2-\sigma^2Q(t)\right]C\st(t) = \sigma^2 Q(t) \avgst{x}^2,
    \label{eq:ODE}
\end{equation}
Here the derivative is with respect to the difference $t-t'$, which we renamed $t$. The boundary conditions of this ODE are $C\st'(0)=0$, since $C(t)$ is even, and $C\st(\infty)=0$, since we expect $x(t)$ to decorrelate with $x(t')$ when $|t-t'|\to\infty$. This ODE is solved with standard methods in \ref{appendix:ODE-solution}.

The solution of the ODE for which the boundary conditions are satisfied has initial condition $C\st(0)$, which is the stationary variance $\sigma\st^2 = \avgst{x^2} - \avgst{x}^2$, given by
\begin{equation}
    \sigma\st^2 = \left[\frac{\hg{1}{2}\left(n;n+1,2n+1;-\lambda^2/4\right)}{2 \hg{0}{1} \left(2n;-\lambda^2/4\right)-\hg{0}{1}\left(2n+1;-\lambda^2/4\right)} - 1\right] \avgst{x}^2,
\label{eq:sigmaeq}
\end{equation}
where
\begin{eqnarray}
    n = k\tau, \label{eq:n} \\
    \lambda = \sqrt{2\tau(1+2\tau)} \sigma \label{eq:lambda}
\end{eqnarray}
and $\hg{p}{q}$ is the generalized hypergeometric function \cite{weisstein2006generalized}. Notice that the dependence on $\mu$ and $h$ in $\sigma\st$ is factorized in the stationary mean $\avgst{x} = h/(k-\mu)$.

\begin{figure}
    \centering
    \includegraphics[width=\textwidth]{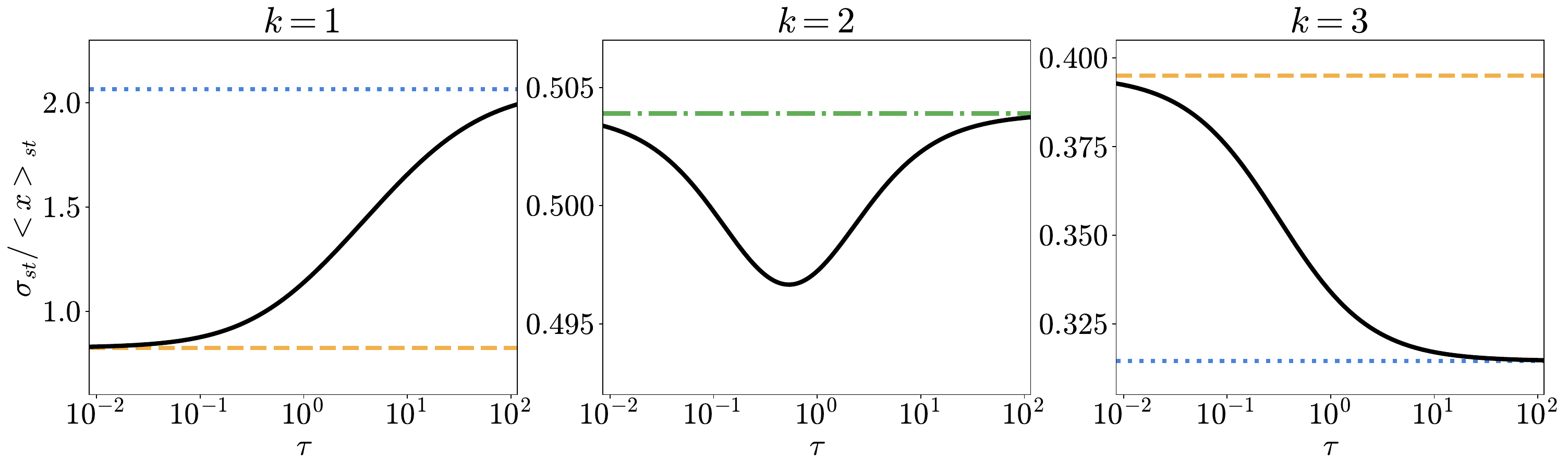}
    \caption{The stationary variance of $x(t)$, given by \eref{eq:sigmaeq}, as a function of $\tau$ for different values of $k$, at fixed $\sigma=0.9$. The dependence of $\sigma\st$ on $\tau$ can also be non-monotonic for other values of $k$, not just for $k=2$ (not shown here). The orange (dashed) is the value of the stationary variance for $\tau\to0$, the blue (dotted) line is the one for $\tau\to\infty$. For $k=2$ the two lines coincide (green, dashed-dotted line).} 
    \label{fig:sigmaeq}
\end{figure}

An example of the dependence of $\sigma\st$ on the parameters of the model is given in \fref{fig:sigmaeq}. Interestingly, $\sigma\st$ can be either a monotonically increasing or decreasing function of $\tau$, depending on the values of $k$ and $\sigma$. For some values of the parameters the dependence of $\sigma\st$ on $\tau$ can even be non-monotonic. Although not shown in \fref{fig:sigmaeq}, we found this non-monotonic dependence of $\sigma\st$ on $\tau$ takes place not only for $k=2$, but also for values of $k$ close to $k=2$.

\Eref{eq:sigmaeq} also allows us to derive the phase diagram of the model, as discussed in the next section.

The explicit expression for $C\st(t)$ is reported in \ref{appendix:ODE-solution}, see \eref{eq:Ceq(t)}. It turns out that $C\st(t) > 0$ at all times for any value of the parameters. Moreover, the characteristic timescale of the autocorrelation is
\begin{equation}
    \tau_c = \frac{\int_0^\infty \rmd t C\st(t)}{C\st(0)}  = 1/k + \tau.
    \label{eq:tauc}
\end{equation}
We verified the last relation by performing the numerical integration in the numerator for different values of $\tau$ while fixing all the other parameters. \Fref{fig:figure5e6} shows comparisons between the analytical and numerical stationary autocorrelation.

\subsection{Power spectral density}
\begin{figure}
    \centering
    \begin{minipage}{0.475\textwidth}
    \centering
    \includegraphics[width=\textwidth]{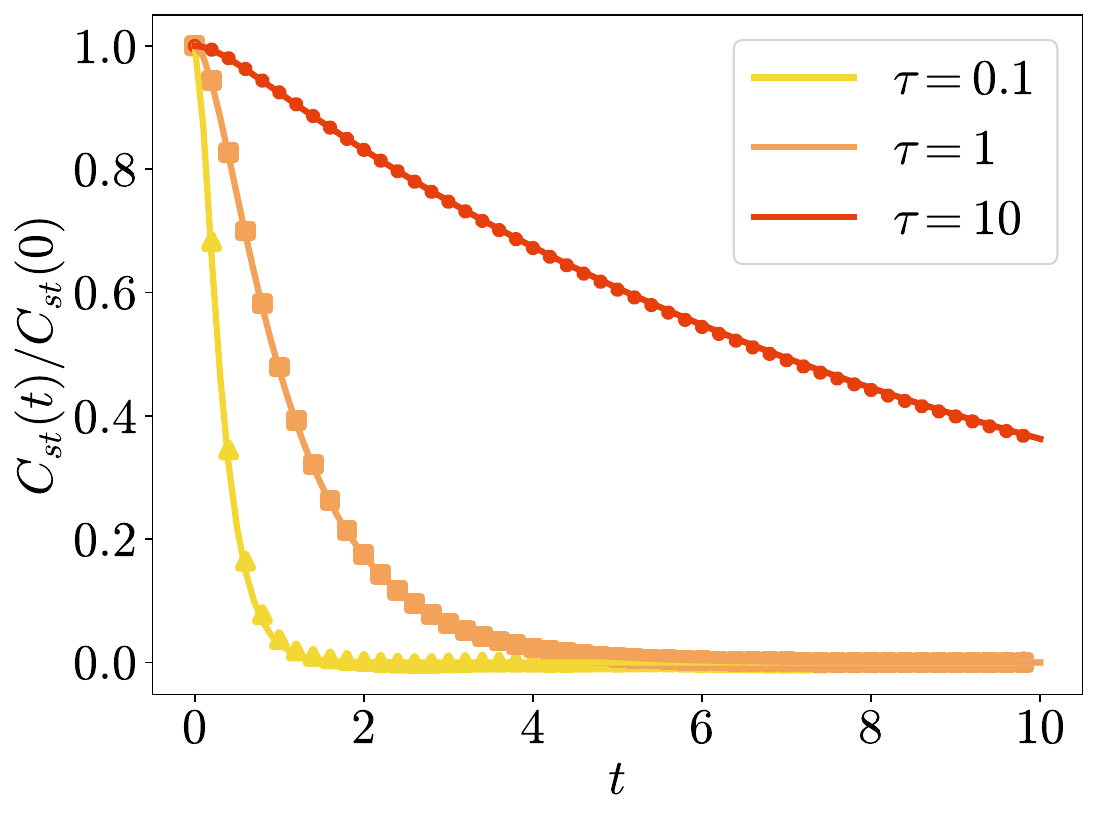}
    \end{minipage}\hfill%
    \begin{minipage}{0.475\textwidth}
    \centering
    \includegraphics[width=\textwidth]{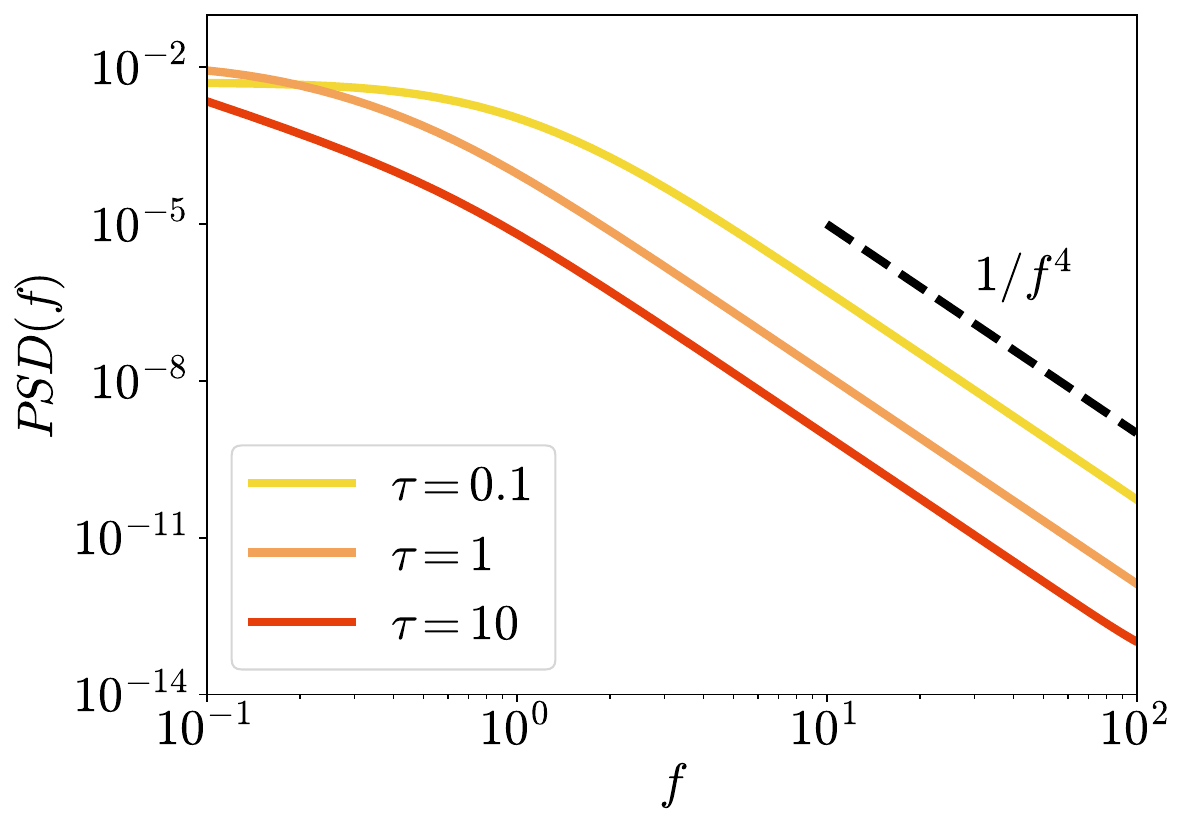}
    \end{minipage}
\caption{Left: Comparison between the analytical result and numerics for the stationary autocorrelation of $x(t)$ for different $\tau$. Solid lines are given by \eref{eq:Ceq(t)}, while markers indicate numerical simulations. The parameters are $N=1000$, $dt=0.01$, $h=1$, $k=4$, $\mu=0$, and $\sigma=1$. Right: Power spectral density of $x(t)$ for different values $\tau$. Lines were obtaining by performing numerically the Fourier transform of \eref{eq:Ceq(t)}. The parameters are $h=1$, $k=4$, $\mu=0$, and $\sigma=1$.}
\label{fig:figure5e6}
\end{figure}

The power spectral density of the process $x(t)$ can be found as the Fourier transform of the stationary autocorrelation. We found no simple expression for the power spectral density, but we verified that it decays as $1/f^4$ at high frequencies by computing numerically the Fourier transform. This is shown in \fref{fig:figure5e6}.

We explain heuristically the exponent of this decay by observing that in the linear equation \eref{eq:linear-equation} the process $x_i(t)$ is essentially the time integration of the Ornstein-Uhlenbeck noise $\alpha_{ij}(t)$, which has a power spectral density that decays as $1/f^2$ at high frequencies. The time integration includes a further factor $1/f^2$, which leads to the decay $1/f^4$.

\section{Phase diagram}\label{sec:phase-diagram}
In this section, we discuss the phases of the linear system as a function of the mean $\mu$ and the variance $\sigma$ of the disorder.

As already discussed in the previous section, for the system to reach the stationary state, we have the necessary condition $\mu<k$.

Furthermore, the stationary variance $\sigma\st^2$ must be finite. By numerical investigation, we observe that increasing $\sigma$ in \eref{eq:sigmaeq}, while keeping $k$ and $\tau$ fixed, a critical $\sigma_c$ is reached such that for $\sigma > \sigma_c$ the stationary variance diverges. This is the smallest $\sigma_c$ for which the denominator in \eref{eq:sigmaeq} vanishes. Thus, we get the following equation for the critical value $\sigma_c$
\begin{equation}
    2 \hg{0}{1} (2 k \tau ;-\tau  (1+2 \tau)\sigma_c ^2/2)-\hg{0}{1}(1+2 k \tau;-\tau  (1+2 \tau)\sigma_c^2/2) = 0.
\end{equation}
This relation can also be simplified using relations between the hypergeometric function and the Bessel functions $J_n(x)$ (equation 9.1.69 of \cite{abramowitz1972handbook})
\begin{equation}
    2n J_{2n}(\lambda_c) - \lambda_c J_{2n-1}(\lambda_c) = 0,
    \label{eq:sigmac-bessel}
\end{equation}
where $n$ and $\lambda_c$ have been defined in \eref{eq:n} and \eref{eq:lambda}, evaluated at the critical $\sigma_c$. Notice that $\sigma_c$ does not depend neither on $h$ nor on $\mu$. Numerical simulations confirm that for $\sigma<\sigma_c$ the system reaches the stationary state, while for $\sigma>\sigma_c$ the variance of the trajectories at fixed time diverges exponentially and a stationary state is not reached.

\begin{figure}
    \centering
    \includegraphics[width=\textwidth]{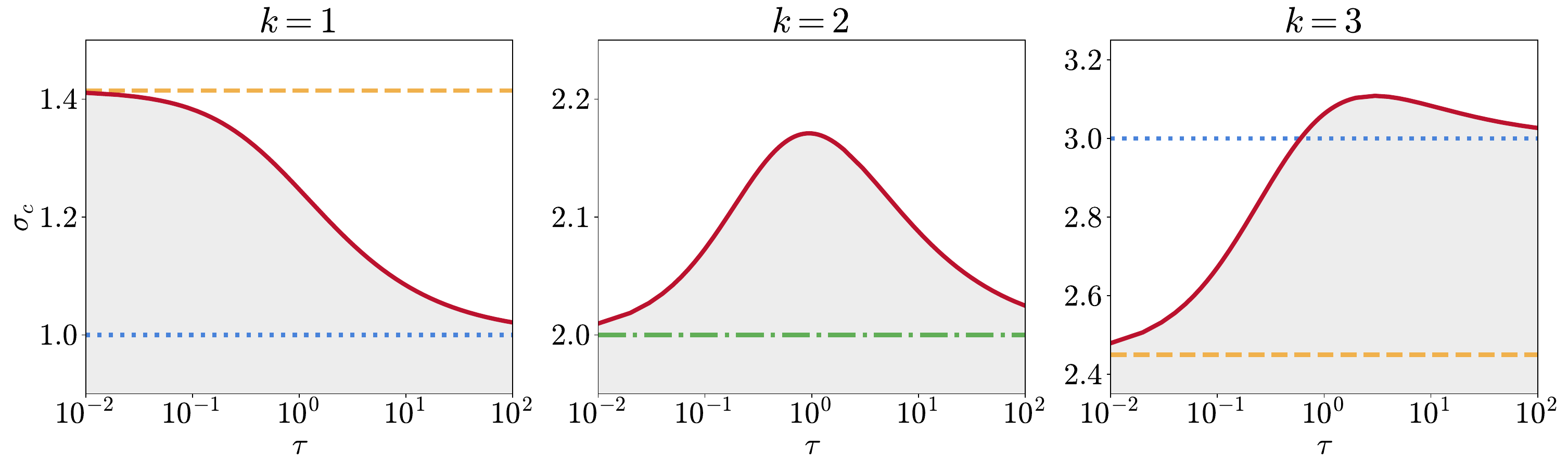}
    \caption{The critical value $\sigma_c$ as a function of $\tau$ for different values of $k$, found as the solution of \eref{eq:sigmac-bessel}. For $\sigma<\sigma_c$ the system reaches a stationary state (darker region). For $\sigma>\sigma_c$ the variance diverges. The horizontal orange (dashed) line is the value of $\sigma_c$ for $\tau\to0$, while the blue (dotted) line is the one for $\tau\to\infty$. For $k=2$ the two lines coincide (green, dashed-dotted line).}
    \label{fig:sigma-critical}
\end{figure}

\Fref{fig:sigma-critical} shows $\sigma_c$ as a function of $\tau$ for different values of $k$. Interestingly, the dependence of $\sigma_c$ on $\tau$ can be both increasing or decreasing. Moreover, it is not necessarily monotonic, displaying in some cases a re-entrant phase transition in $\tau$. This means that at fixed $\sigma$, increasing $\tau$ can sometimes move the system from a phase in which the stationary state is not reached, to a phase in which it is reached, and finally to a phase where it is not reached again.

We also notice that for $k<2$ the critical $\sigma_c$ at $\tau=0$ is larger than the critical $\sigma_c$ at $\tau=\infty$, while for $k>2$ the opposite is true. At $k=2$ the two values coincide, but the dependence of $\sigma_c$ on $\tau$ is nevertheless non-trivial.

\begin{figure}
    \centering
    \includegraphics[width=0.7\textwidth]{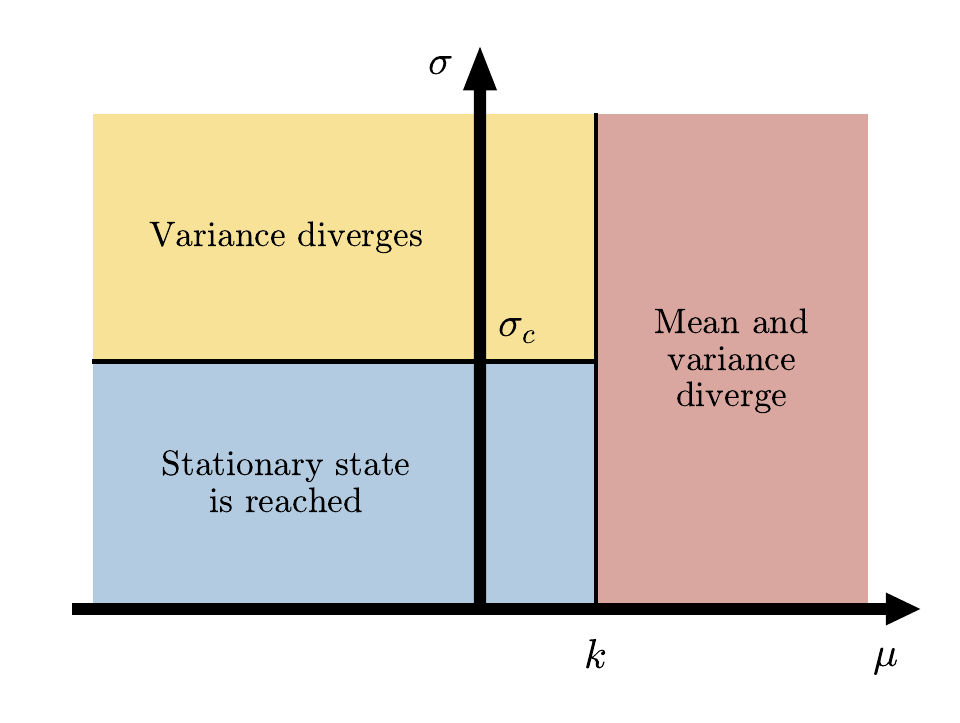}
    \caption{The phase diagram of the model as a function of the mean and the variance of the annealed disorder. For $\mu<k$ and $\sigma < \sigma_c$ the system reaches a stationary state, in which the stationary distribution is a Gaussian. For $\mu<k$ and $\sigma>\sigma_c$ this Gaussian has a finite mean but its variance diverges. For $\mu>k$ both the mean and the variance of this Gaussian diverge.}
    \label{fig:phase-diagram}
\end{figure}

In conclusion, the phase diagram of the system is the one reported in \fref{fig:phase-diagram}.

\section{White-noise and quenched limits}\label{sec:white-quenched}
The previous results simplify in the limits $\tau\to0$ and $\tau\to\infty$.

\subsection{White-noise limit, $\tau\to0$}
In the white-noise limit, the noise $\eta(t)$ in the DMFT equation \eref{eq:DMFT} becomes white, although with an amplitude which is time-dependent
\begin{equation}
    \avg{\eta(t)\eta(t')} = \avg{x(t)^2} \delta(t-t').
\end{equation}
In this limit a simple expression for the equal-time average of the noise $\eta(t)$ and the process $x(t)$ can be easily found
\begin{equation}
    \avg{x(t)\eta(t)} = \frac12 \sigma \avg{x(t)^2},
\end{equation}
which allows us to write a simple equation for the evolution of the variance $s^2(t) = \avg{x(t)^2} - \avg{x(t)}^2$
\begin{equation}
    \frac{\rmd s^2(t)}{\rmd t} = \sigma^2 \avg{x(t)}^2 - (2k-\sigma^2)s^2(t).
\end{equation}
From this equation, we read explicitly that
\begin{equation}
    \sigma_c = \sqrt{2k}
\end{equation}
and, when stationarity is reached, the variance is 
\begin{equation}
    \sigma^2\st = \frac{\sigma^2}{2k-\sigma^2} \avgst{x}^2.
\end{equation}
This result can also be derived by taking the $\tau\to0$ limit of \eref{eq:sigmaeq}. The stationary auto-correlation is also simply
\begin{equation}
    C\st(t) = \sigma^2\st \exp(-k\abs{t}).
\end{equation}

\subsection{Quenched disorder limit, $\tau\to\infty$}
In the quenched disorder limit the mean and the correlation of the noise in the DMFT equation \eref{eq:DMFT} reads
\begin{equation}
    \avg{\eta(t)\eta(t')} = \avg{x(t)x(t')}.
\end{equation}
The stationary variance can be obtained either by taking the limit $\tau\to\infty$ of \eref{eq:sigmaeq} or by a fixed-point ansatz in the DMFT equation \eref{eq:DMFT}. It is given by
\begin{equation}
    \sigma^2\st = \frac{\sigma^2}{k^2-\sigma^2} \avgst{x}^2,
\end{equation}
so that the critical $\sigma$ is simply
\begin{equation}
    \sigma_c = k.
\end{equation}
This critical $\sigma_c$ could also be obtained by applying the circular law \cite{girko1985circular}.

\section{Unified Colored-Noise Approximation}\label{sec:UCNA}
In the previous sections, we have shown that the presence of annealed disorder makes it challenging to find exact solutions, even in the simplest case of a linear equation. This casts doubt on the possibility of solving exactly more complicated models with annealed disorder. This is because the DMFT equation is driven by a noise that is colored and self-consistent. Currently, there exists no general theory to describe non-linear systems driven by colored noise \cite{haunggi1994colored}, let alone one addressing self-consistent noise. In this section, we thus present an approximation scheme that is applicable to any model with annealed disorder.

To deal with colored noise, a commonly used approximation scheme is the Unified Colored Noise Approximation (UCNA) \cite{jung1987dynamical}. UCNA gives an approximate analytical expression for the stationary state of a general one-dimensional Langevin equation driven by exponentially correlated Gaussian noise, which is exact in both the limits $\tau\to0$ and $\tau\to\infty$. This approximation is based on a rescaling of the time variable which makes an adiabatic elimination procedure possible.

To take into account the self-consistency relation, we start by assuming that the stationary autocorrelation decays exponentially with a characteristic time $\tau_x$
\begin{equation}
    \avgst{x(t) x(t')} \approx \left( \avgst{x^2} -\avgst{x}^2\right) \rme^{-|t-t'|/\tau_x}+ \avgst{x} ^2.
\end{equation}
For $|t-t'| \ll \tau_x$, we can expand the previous equation as
\begin{equation}
     \avgst{x(t) x(t')} \approx \avgst{x^2} - \left(\avgst{x^2}- \avgst{x}^2\right) \frac{|t-t'|}{\tau_x} \approx \avgst{x^2} \rme^{-|t-t'|/\tau'_x},
\end{equation}
with
\begin{equation}
    \tau'_x = \frac{\avgst{x^2}}{\avgst{x^2}- \avgst{x}^2} \tau_x.
\end{equation} Thus the autocorrelation of the noise $\eta(t)$ at stationarity can be approximated as
\begin{equation}
        \avgst{\eta(t) \eta(t')} \approx \avgst{x^2} \frac{1+2\tau}{2 \tau}\rme^{- |\Delta t|/\overline{\tau}},\label{eq:autocorrelation_noise}
\end{equation}
where the correlation time is
\begin{equation}
    \overline{\tau} = \left( \frac{1}{\tau} + \frac{1}{\tau'_x} \right) ^{-1}.
    \label{eq:tau-bar}
\end{equation}

With these approximations, we can now use the UCNA. Following the notation of \cite{jung1987dynamical}, with $f(x) = h - k x + \mu \avgst{x}$ and $g(x) = \sigma$, we obtain the stationary distribution
\begin{equation}
    P\st(x) = \frac{1}{\sqrt{2 \pi\sigma\st^2} } \exp\left[ - \frac{(x- \avgst{x})^2}{2 \sigma\st^2}\right],\label{eq:Pst-UCNA}
\end{equation}
where the mean and the variance of the distribution are given by
\begin{eqnarray}
    \avgst{x} = \frac{h}{k-\mu}, \\
    \sigma\st^2 = \frac{\sigma^2 \overline{\tau} (1+2\tau)}{2 \tau (k+k^2 \overline{\tau}) - (1+2 \tau) \overline{\tau} \sigma^2}  \avgst{x}^2 \label{eq:variance-UCNA}.
\end{eqnarray}

\begin{figure}
    \centering
    \includegraphics[width=\textwidth]{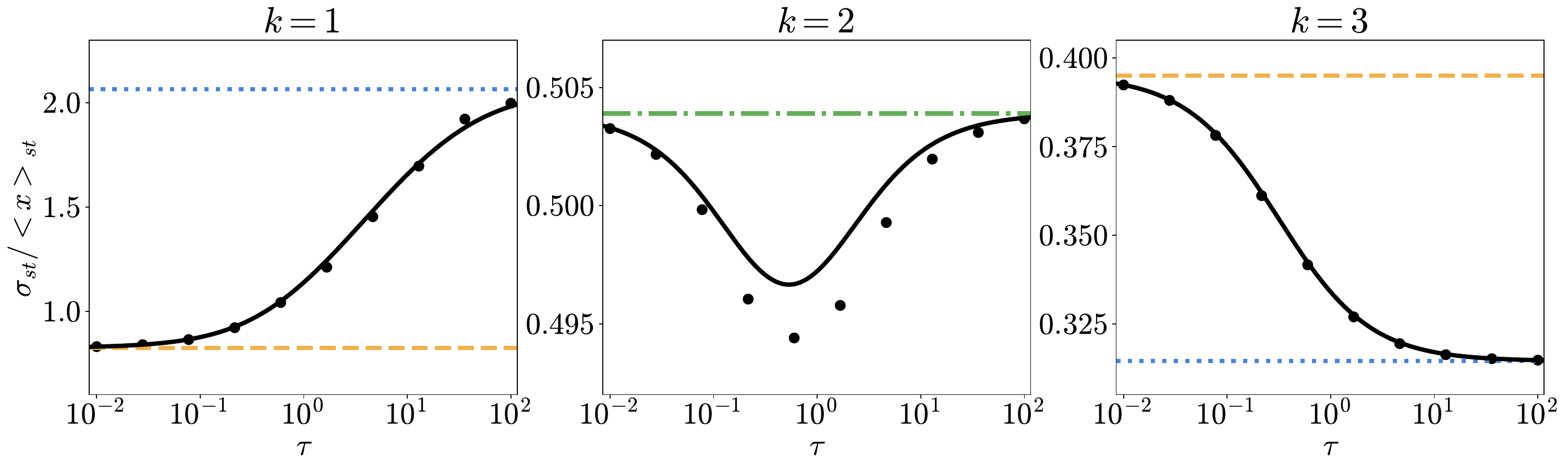}
    \caption{The stationary variance $\sigma\st^{2}$ approximated using UCNA as a function of $\tau$ for different values of $k$, at fixed $\sigma=0.9$. The orange (dashed) is the value of the stationary variance for $\tau\to0$, the blue (dotted) line is the one for $\tau\to\infty$. The solid black line is the exact stationary variance \eref{eq:sigmaeq}, while the markers indicate the variance obtained approximated using UCNA. The parameters of the simulation are $N=1000$, $dt=0.001$, $h=1$, and $\mu=0$.}
    \label{fig:variance-UCNA}
\end{figure}

Our approximation scheme is thus able to recover the exact Gaussian stationary distribution, although with an approximate variance. The stationary variance \eref{eq:variance-UCNA} depends on the unknown parameter $\overline{\tau}$, which can be found by fitting the autocorrelation of the noise \eref{eq:autocorrelation_noise}. \Fref{fig:variance-UCNA} shows that our approximation scheme is in very good agreement with the exact solution at finite $\tau$, and that the two results coincide at $\tau \rightarrow 0$ and $\tau \rightarrow \infty$.

\section{Numerics}
\label{sec:numerics}
All numerical simulations reported in this work have been performed by integrating the linear system \eref{eq:linear-equation} using the second-order Heun's method \cite{ricardo2020modern}. This algorithm was chosen for an easier comparison of the results with the white-noise limit, where the Stratonovich-Heun algorithm \cite{burrage2004numerical} was used. The Ornstein-Uhlenbeck noise was generated using the Bartosch method \cite{bartosch2001generation}.

\section{Summary and outlook}

In this work, we have studied a disordered linear system in which interactions are not fixed in time but vary stochastically with a correlation time $\tau$. We dubbed this type of disorder annealed disorder. By employing Dynamical Mean-Field Theory, tailored for the case of annealed disorder, we were able to find the exact solution of the system in the limit of a large number of degrees of freedom.

Dynamical Mean-Field Theory leads to an equation similar to the stochastic differential equation of an Ornstein-Uhlenbeck process, with the crucial difference that it is driven by self-consistent and colored noise. This difference makes analytical results much more challenging and gives rise to non-trivial behaviors.

We derived equations for the mean and autocorrelation of the Dynamical Mean-Field Theory process and solved them. Since this process is Gaussian, this solves exactly the model.

The exact solution allowed us to derive the phase diagram of the linear system. We found that if the mean or the variance of the interactions exceeds a critical threshold, the system does not reach a stationary state. Interestingly, our findings reveal that the critical variance value $\sigma_c$ depends in a non-trivial way on the correlation time $\tau$ of the annealed disorder. Furthermore, when the stationary state is reached, we have shown that its variance also varies in a non-monotonic way depending on $\tau$.

The solution of the model allowed us to compare the exact stationary state with the one obtained employing an extension of the Unified Colored Noise Approximation. This approximation is able to recover the correct Gaussian functional form of the stationary distribution. We also showed that it recovers the exact mean and gives an approximated variance very close to the exact one for any value of $\tau$.

We foresee a number of possible extensions with respect to the linear model studied in this work. Firstly, we assumed that the interaction parameters $\alpha_{ij}(t)$ were independent of each other, but a correlation between pairs of couplings \cite{bunin2017ecological,galla2018dynamically} or a hierarchical structure \cite{poley2023generalized} could be introduced. For instance, human microbiomes exhibit both taxonomic and functional organization far more intricate than the corresponding null models consisting of entirely uncorrelated species or functions \cite{seppi2023emergent,camacho2024sparse}. Moreover, we assumed the model to be fully connected, but, as has been shown recently \cite{park2024incorporating,poley2024interaction,aguirre2024heterogeneous}, the Dynamical Mean-Field Theory approach can also be applied to situations in which a network structure is present in the interactions. It would also be interesting to investigate the effect of non-Gaussian interactions on the emerging properties of these systems \cite{azaele2023large}.

More broadly, the framework of annealed disorder considered in this work could be applied to any many-body system in which interactions can be modeled as random, but not static. Investigating the impact of annealed disorder on the properties of the diverse number of models that have been treated so far in the limit of quenched disorder would be a compelling research avenue. Eventually, exploring how a combination of quenched and annealed disorder may produce emergent patterns observed in physical or biological systems could yield valuable insights to bridge the gap between theoretical models and real-world systems.

\ack

We thank very much Tommaso Jack Leonardi for a careful reading of the manuscript. F.F. was supported by the Italian Ministry of University and Research (project funded by the European  Union - Next Generation EU: ``PNRR Missione 4 Componente 2, ``Dalla ricerca all’impresa'', Investimento 1.4, Progetto  CN00000033''). C.G., A.M. and S.A. acknowledge financial support under the National Recovery and Resilience Plan (NRRP), Mission 4, Component 2, Investment 1.1, Call for tender No. 104 published on 2.2.2022 by the Italian Ministry of University and Research (MUR), funded by the European Union – NextGenerationEU – Project Title ``Emergent Dynamical Patterns of Disordered Systems with Applications to Natural Communities'' – CUP 2022WPHMXK - Grant Assignment Decree No. 2022WPHMXK adopted on 19/09/2023 by the Italian Ministry of Ministry of University and Research (MUR). S.S. acknowledges financial support under the National Recovery and Resilience Plan (NRRP), Mission 4, Component 2, Investment 1.1, Call for tender No. 104 published on 2.2.2022 by the Italian Ministry of University and Research (MUR), funded by the European Union – NextGenerationEU – Project Title: Anchialos: diversity, function, and resilience of Italian coastal aquifers upon global climatic changes – CUP C53D23003420001 Grant Assignment Decree n. 1015 adopted on 07/07/2023 by the Italian Ministry of Ministry of University and Research (MUR).

\appendix

\section{Derivation of the DMFT equation}
\label{appendix:DMFT_derivation}

In this Appendix we derive the DMFT equation for the linear system \eref{eq:linear-equation}. The derivation is similar to the one presented in \cite{suweis2023generalized}, which is in turn a simple extension of the one given in \cite{galla2018dynamically}.

We start by considering the generating functional of \eref{eq:linear-equation}, which defined as
 \begin{equation}
     Z[\psi] = \int D[x] \delta(x-x^*) \rme^{i \psi \cdot x},
 \end{equation}
where $x_i^*(t)$ are solutions of \eref{eq:linear-equation} for a given realization of the disorder and $\psi_i(t)$ are external source fields that generate correlation functions and will eventually be set to zero. We introduced the shorthand notation $\psi \cdot x= \sum_i \int dt \psi_i(t) x_i(t)$.

Assuming the system to be self-averaging, we average the generating functional over the disorder 
 \begin{equation}
     \overline{Z[\psi]} = \int D[x] \overline{\delta(x-x^*)} \rme^{i \psi \cdot x} = \int D[x] \overline{\delta(E[x])} \rme^{i \psi \cdot x},
 \end{equation}
where we changed the argument of the delta-function to the equation of motion
 \begin{equation}
     E[x_i(t)] = \dot{x}_i(t) - \left[h - k x_i(t) + \sum_{j \neq i} \alpha_{i j}(t) x_j(t) \right].
 \end{equation}
To perform the average over the disorder we introduce the conjugate variables $\hat{x}_i(t)$ to represent the delta-function as a Fourier transform
\begin{equation}
    \overline{Z[\psi]} = \int D[x, \hat{x}] \overline{\exp{\left( i \sum_i \int dt \hat{x}_i(t) E[x_i(t)] \right)}} \rme^{i \psi \cdot x},
    \label{eq:generating-functional}
\end{equation}
The only term to be averaged over the disorder can be computed as a Gaussian integral
\begin{equation}
\eqalign{
\fl
      \overline{\exp \left( -i \sum_{ij} \int dt \hat{x}_i (t) \alpha_{ij} (t) x_j(t)  \right)} &= \exp \left[ -N \mu \int dt M(t) N(t)  \right]\\
      &\quad\times \exp \left[ -\frac{1}{2} N \sigma^2 \int dt dt' Q(t-t') C(t,t') D(t,t') \right],\label{eq:disorder-average}
      }
 \end{equation}
where we have introduced the order parameters
\begin{eqnarray}
    M(t) = \frac{1}{N} \sum_{i} x_i(t),\\
    N(t) = i \frac{1}{N} \sum_{i} \hat{x}_i(t),\\
    C(t,t') = \frac{1}{N} \sum_{i} x_i(t) x_i(t'),\\
    D(t,t') = \frac{1}{N} \sum_{i} \hat{x}_i(t) \hat{x}_i(t').
\end{eqnarray}
In taking the thermodynamic limit it is convenient to introduce the order parameters and their corresponding conjugate as delta-function, as for instance
\begin{eqnarray}
\eqalign{
    1 &= \int D[M] \delta \left( M(t) - \frac{1}{N} \sum_{i} x_i(t)   \right)\\
    &= \int D[M,\hat{M}] \exp{\left[iN \int dt \hat{M}(t) \left( M(t) -\frac{1}{N} \sum_{i} x_i(t) \right) \right]}.
    }
\end{eqnarray}

The disorder-averaged generating functional is then, after some manipulations,
\begin{equation}
    \overline{Z[\psi]} = \int \rme^{S(\Psi + \Phi + \Omega)},\label{eq:saddle-point}
\end{equation}
where the integral is over the hatted and non-hatted order parameters. The first term in the exponent results from the introduction of the order parameters
\begin{eqnarray}
\eqalign{
    \Psi &=  i\int dt \left(\hat{M}(t)M(t) + \hat{N}(t)N(t)\right)\\
    &\quad+ i\int dt dt' \left(\hat{C}(t,t')C(t,t') + \hat{D}(t,t')D(t,t')\right),
    }
\end{eqnarray}
the second term results from the average over the disorder
\begin{equation}
    \Phi =
    -\mu\int dt M(t)N(t)
    -\frac12\sigma^2 \int dtdt' Q(t-t')C(t,t')D(t,t'),
\end{equation}
and the last term contains all the information on the microscopic dynamics 
\begin{eqnarray}
    \Omega = \frac{1}{N} \sum_i \log Z_i, \\
    Z_i = \int D[x_i, \hat{x}_i] \exp{\left[ i \Omega_i \right]}
\end{eqnarray}
\begin{eqnarray}
\eqalign{
    \Omega_i &= \int dt \hat{x}_i(t)\left[\dot{x}_i(t)-(h_i(t) - k x_i(t) )\right]
    + \int dt \psi_i(t) x_i(t) \\
     &\quad-\int dt \left(\hat{M}(t)x_i(t)+i\hat{N}(t)\hat{x}_i(t)\right)\\
     &\quad-\int dtdt'\left(\hat{C}(t,t')x_i(t)x_i(t') + \hat{D}(t,t')\hat{x}_i(t)\hat{x}_i(t')\right).\label{eq:omega_i}
     }
\end{eqnarray}

We now use the saddle-point approximation to evaluate the integral in \eref{eq:saddle-point}. Extremizing the exponent with respect to the non-hatted order parameters gives 
\begin{eqnarray}
    i\hat{M}(t) = \mu N(t), \\
    i\hat{N}(t) = \mu M(t), \\
    i\hat{C}(t,t') = \frac12\sigma^2 Q(t-t') D(t,t'), \\
    i\hat{D}(t,t') = \frac12\sigma^2 Q(t-t') C(t,t'),
\label{eq:saddle-point-non-hatted}
\end{eqnarray}
while extremising it with respect to the hatted variables gives, in the thermodynamic limit,
\begin{eqnarray}
    M(t) = \lim_{N\to\infty} \frac{1}{N} \sum_{i=1}^N \avg{x_i(t)}_\Omega, \\ 
    N(t) = \lim_{N\to\infty} \frac{1}{N} \sum_{i=1}^N i \avg{\hat{x}_i(t)}_\Omega, \\ 
    C(t,t') = \lim_{N\to\infty} \frac{1}{N} \sum_{i=1}^N\avg{x_i(t)x_i(t')}_\Omega, \\ 
    D(t,t') = \lim_{N\to\infty} \frac{1}{N} \sum_{i=1}^N \avg{\hat{x}_i(t)\hat{x}_i(t')}_\Omega,
\end{eqnarray}
where $\avg{...}_\Omega$ is the average taken with the action defined in equation \eref{eq:omega_i}. It turns out \cite{galla2018dynamically,suweis2023generalized} that at the saddle-point $N(t)=0$, $D(t,t')=0$, $\hat{M}(t)=0$, and $\hat{C}(t,t')=0$.

We now set $\psi=0$. Simple manipulations show that the disorder-averaged generating functional evaluated in the thermodynamic limit reduces to a non-interacting problem
\begin{equation}
    \overline{Z} = Z_\mathrm{eff}^N
\end{equation}
where 
\begin{eqnarray}
\eqalign{
    Z_\mathrm{eff}[\psi] = \int &D[x, \hat{x}]\exp{\left[i \int dt \hat{x}(t) \left( \dot{x}(t) - \left( h - k x(t) +\mu M(t) \right)  \right)   \right]} \\
&\times \exp{\left[ -\frac{1}{2} \sigma^2 \int dt dt' Q(t-t') C(t,t') \hat{x}(t) \hat{x}(t')\right]},
}
\end{eqnarray}
We can rewrite the last square bracket  introducing the Gaussian variable $\eta (t)$,
\begin{eqnarray}
\eqalign{
\fl
    \exp\bigg[
        -\frac12&\sigma^2\int dt dt' Q(t-t')C(t,t') \hat{x}(t) \hat{x}(t')
    \bigg] \\
\fl
    &= \frac{1}{Z_\eta}
    \int D[\eta]
    \exp\left[
        -i \sigma \int dt \eta(t) \hat{x}(t)
    \right]
    \exp\left[
        -\frac{1}{2} \int dtdt'  \eta(t) A(t,t')\eta(t')
    \right]
}
\end{eqnarray}
where $A(t,t')$ is the inverse of $Q(t-t')C(t,t')$.

It finally follows that $Z_\mathrm{eff}$ is the generating functional of \eref{eq:DMFT}, where self-consistently
\begin{eqnarray}
    M(t) = \avg{x(t)},\\
    \avg{\eta(t) \eta(t')} = Q(t-t') \avg{x(t) x(t')}.
\end{eqnarray}

\section{Solution of PDE for autocorrelation}
\label{appendix:PDE-solution}

In this Appendix we solve the PDE \eref{eq:PDE}.

Consider the transformation $C(t,t') = \rme^{-k(t+t')}D(t,t')$. The equation for $D(t,t')$ is the simpler PDE
\begin{equation}
    \left[\partial_t\partial_{t'}-\sigma^2Q(t-t')\right] D(t,t') = \rme^{k(t+t')}f(t,t'),
    \label{eq:PDE-D}
\end{equation}
together with the boundary conditions $D(t,0)=0$ and $D(0,t')=0$. To solve this inhomogeneous PDE we use the Riemann method \cite{garabedian2023partial}. In the present case it amounts to finding the Riemann function $A(s,s';t,t')$, which is defined as the solution of the homogeneous PDE
\begin{equation}
    \left[\partial_s\partial_{s'}-\sigma^2Q(s-s')\right] A(s,s';t,t') = 0,
    \label{eq:Riemann-PDE}
\end{equation}
together with the boundary conditions $A(s,t';t,t')=1$ and $A(t,s';t,t')=1$. Call $A_0(s,s';t,t')$ the Riemann function of the PDE were the absolute value function not present in $Q(s-s')$. The PDE for $A_0$ could be solved by performing the change of variables $u=\rme^{-s/\tau}$, $u'=\rme^{s'/\tau}$, and using the known solution \cite{garabedian2023partial} of the equation $(\partial_u\partial_{u'}+c)A_0(u,u')=0$ to get
\begin{equation}
    A_0(s,s';t,t') = J_0\left(\lambda\sqrt{(\rme^{-s/\tau}-\rme^{-t/\tau})(\rme^{s'/\tau}-\rme^{t'/\tau})}\right)
\end{equation}
where $\lambda=\sqrt{2\tau(1+2\tau)}\sigma$. Similarly, were the absolute value in $Q(s-s')$ to be replaced with the opposite of its argument, the solution of \eref{eq:Riemann-PDE} would be $A_0(-s,-s';-t,-t')$. It is then clear that the Riemann function in the presence of the absolute value in $Q(s-s')$ is
\begin{equation}
    A(s,s';t,t') = \theta(s-s') A_0(s,s';t,t') + \theta(s'-s)A_0(-s,-s';-t,-t').
    \label{eq:Riemann-function}
\end{equation}
Using the Riemann formula the solution of \eref{eq:PDE-D} is
\begin{equation}
    D(t,t') = \int_0^t \rmd s \int_0^{t'} \rmd s' \rme^{k(s+s')} A(s,s';t,t') f(s,s').
\end{equation}
In terms of $C(t,t')$ one immediately concludes that the solution of \eref{eq:PDE} is \eref{eq:PDE-solution}.

\section{Solution of ODE}
\label{appendix:ODE-solution}
In this Appendix we solve the ODE \eref{eq:ODE}.

It is simpler to solve for the function $E(t) = \avgst{x(t)x(0)} = C\st(t) + \avgst{x}^2$. The ODE for $E(t)$ is
\begin{equation}
    -E''(t) + \left[k^2-\sigma^2Q(t)\right] E(t) = k^2 \avgst{x}^2.
    \label{eq:ODE-E}
\end{equation}
With the change of variable $z=\lambda \rme^{-t/2\tau}$, $\lambda = \sqrt{2\tau(1+2\tau)}\sigma$ the homogeneous equation becomes
\begin{equation}
    z^2 E''(z) + z E'(z) + \left[z^2 - (2k\tau)^2\right] E(z) = 0,
\end{equation}
which is Bessel differential equation. Two independent solutions are then
\begin{eqnarray}
    u_1(t) &= J_{2k\tau}(\lambda \rme^{-t/2\tau}), \\
    u_2(t) &= Y_{2k\tau}(\lambda \rme^{-t/2\tau}).
\end{eqnarray}
Since the Wronskian of $u_1(t)$ and $u_2(t)$ is $W=1/\pi\tau$, the general solution of \eref{eq:ODE-E} is, by the method of variation of parameters,
\begin{equation}
    E(t) = c_1 u_1(t) + c_2 u_2(t) + \pi\tau k^2 \avgst{x}^2 \left[ u_2(t) U_1(t) - u_1(t) U_2(t)\right],
    \label{eq:ODE-E-solution}
\end{equation}
where
\begin{equation}
\fl
    U_1(t) = \int \rmd t\, u_1(t) = - \frac{(\lambda/2)^{2k\tau}}{k\Gamma(1+2k\tau)} \rme^{-kt} \hg{1}{2}(k\tau;1+2k\tau,1+k\tau;-(\lambda/2)^2 \rme^{-t/\tau})
\end{equation}
and
\begin{equation}
\eqalign{
\fl
    U_2(t) &= \int \rmd t\, u_2(t)  \\
\fl 
    &=\frac{(\lambda/2)^{2k\tau}\cos(2\pi k\tau)\Gamma(-2k\tau)}{\pi k}\rme^{-kt} \hg{1}{2}(k\tau;1+2k\tau,1+k\tau;-(\lambda/2)^2 \rme^{-t/\tau}) \\
\fl
    &\qquad-\frac{(\lambda/2)^{-2k\tau}\Gamma(2k\tau)}{\pi k}\rme^{kt} \hg{1}{2}(-k\tau;1-2k\tau,1-k\tau;-(\lambda/2)^2 \rme^{-t/\tau}).
}
\end{equation}
The primitive functions were found with the substitution $z=\lambda \rme^{-t/2\tau}$ and using known integrals of Bessel functions.

We now impose the boundary conditions. The asymptotic behaviour of each piece of \eref{eq:ODE-E-solution} at long time is
\begin{eqnarray}
    u_1(t) \sim \frac{(\lambda/2)^{2k\tau}}{\Gamma(1+2k\tau)} \rme^{-kt}, \\
    u_2(t) \sim -\frac{\Gamma(2k\tau)(\lambda/2)^{-2k\tau}}{\pi} \rme^{kt}, \\
    u_1(t)U_2(t) \sim -\frac{1}{2\pi k^2 \tau}, \\
    u_2(t) U_1(t) \sim \frac{1}{2\pi k^2 \tau},
\end{eqnarray}
from which we get, if we want $E(\infty)$ to be finite, that $c_2=0$. With this condition we get also $E(\infty)=\avgst{x}^2$. Imposing $E'(0)=0$ we get the condition
\begin{equation}
    c_1 = -\frac12\avgst{x}^2(a+b),
\end{equation}
with
\begin{eqnarray}
    a = (\lambda/2)^{-2k\tau} \Gamma(1+2k\tau) \hg{1}{2}(-k\tau;1-2k\tau,1-k\tau;-(\lambda/2)^2), \\
    \eqalign{
        b &= (\lambda/2)^{2k\tau}\Gamma(1-2k\tau) \frac{k J_{-2k\tau}(\lambda) + \lambda J_{-2k\tau+1}(\lambda)/2\tau}{k J_{2k\tau}(\lambda) - \lambda J_{2k\tau-1}(\lambda)/2\tau} \\
        &\qquad\times\hg{1}{2}(k\tau;1+k\tau,1+2k\tau;-(\lambda/2)^2).
    }
\end{eqnarray}

With all this, the solution of \eref{eq:ODE} is
\begin{equation}
    C\st(t) = c_1 u_1(t) + \pi\tau k^2 \avgst{x}^2 \left[u_2(t) U_1(t) - u_1(t) U_2(t)\right] - \avgst{x}^2.
\label{eq:Ceq(t)}
\end{equation}
Setting $t=0$ and after some manipulations (done with FullSimplify on Mathematica 13.2) one gets \eref{eq:sigmaeq} of the main text.

\section*{References}\par
\bibliographystyle{unsrt}
\bibliography{refs}

\end{document}